\documentclass[a4paper,twoside]{article}
\newif\ifdraft \global\drafttrue
\def\production{\global\draftfalse}
\production
\usepackage[T1]{fontenc}
\usepackage[latin1]{inputenc}
\usepackage{a4wide}
\usepackage{times}
\usepackage{graphicx}
\usepackage{theorem}
\usepackage[colorlinks=true,hyperindex=true]{hyperref}
\usepackage{color}
\usepackage{fancyhdr}
\usepackage{latexsym}
\usepackage{amsmath}
\usepackage{bbm}
\usepackage{amssymb}
\usepackage{amsfonts}
\usepackage{enumerate}
\usepackage{cancel}
\ifdraft
\usepackage{showkeys}
\fi
\newcounter{smallarabics}

\newcounter{smallroman}

\newcommand{\ben}{\begin{enumerate}[{\rm (1)}]}
\newcommand{\een}{\end{enumerate}}

\newtheorem{theoreme}{Theorem }[section]
\newtheorem{proposition}[theoreme]{Proposition}
\newtheorem{lemma}[theoreme]{Lemma}
\newtheorem{definition}[theoreme]{Definition}
\newtheorem{corollary}[theoreme]{Corollary}

\def\rr{{\mathbb R}}
\def\zz{{\mathbb Z}}
\def\cc{{\mathbb C}}

\def\Z{{\mathbb Z}}

\def\textsl{{}}

\def\Im{{\rm Im}\,}
\def\Re{{\rm Re}\,}
\def\ch{\mathfrak{h}}

\newcommand{\slim}{\mathop{\mathrm{s-lim}}\limits}

\def\c0inf{C_0^\infty}
\def\bep{\begin{proposition}}
\def\eep{\end{proposition}}

\def\proof{\noindent {\bf Proof.}\ \ }

\def\E{{\cal  E}}

\def\i{{\rm i}}
\newcommand{\beq}{\begin{equation}}
\newcommand{\eeq}{\end{equation}}
\newcommand{\bear}[1]{\begin{array}{#1}}
\newcommand{\ear}{\end{array}}

\def\sp{{\hat e}}

\newcommand{\e}{\mathrm{e}}
\renewcommand{\i}{\mathrm{i}}

\renewcommand{\d}{\mathrm{d}}

\def\qed{$\Box$\medskip}

\def\bel{\begin{lemma}}
\def\eel{\end{lemma}}
\def\bet{\begin{theoreme}}
\def\eet{\end{theoreme}}
\def\bed{\begin{definition}}
\def\eed{\end{definition}}
\def\bar{\overline}

\def\12{\frac{1}{2}}

\def\e{{\rm e}}

\def\d{{\rm d}}
\def\Ran{{\rm Ran}\,}

\def\one{{\mathbbm 1}}

\def\ac{{\rm ac}}

\def\sp{{\rm sp}}
\def\cS{{\cal S}}

\def\fh{\mathfrak{h}}

\def\dto{{\downarrow}}
\def\eps{\varepsilon}
\def\ee{\mathfrak{e}}
\def\fh{\mathcal{H}}
\def\fhln{\mathcal{H}^{(l)}_n}
\def\fhrn{\mathcal{H}^{(r)}_n}
\def\Jln{J^{(l)}_n}
\def\Jrn{J^{(r)}_n}
\def\mln{m^{(l)}_n}
\def\mrn{m^{(r)}_n}
\def\mlrn{m^{(l/r)}_n}
\def\Jl0{J^{(l)}_{0}}
\def\Jr0{J^{(r)}_0}
\def\dnl{\d \nu_{l, \ac} }
\def\dnr{\d \nu_{r, \ac} }
\def\dl0{\delta_{-1}}
\def\dr0{\delta_{1}}
\def\mlnp{m^{(l)}_{n+1}}
\def\mrnm{m^{(r)}_{n-1}}
\def\cln{\chi^{(l)}_n}
\def\crn{\chi^{(r)}_n}
\def\chl0{\chi^{(l)}_0}
\def\chr0{\chi^{(r)}_0}
\def\ch0lr{\chi^{(l/r)}_0}
\def\de0{\delta_0}
\def\dlr{\delta_{-1 / 1 }}
\def\Jlr0{J^{(l/r)}_0}
\def\dnlr{\mathrm{d} \nu_{l/r, \ac} }
\def\dnj{\mathrm{d} \nu_{j, \ac} }
\def\dnk{\mathrm{d} \nu_{k, \ac} }

\def\mfh{\mathfrak h}
\begin{document}
\def\today{}
\title{A note on  reflectionless  Jacobi matrices}
\author{V. Jak\v{s}i\'c$^{1}$,  B. Landon$^{2}$,  A. Panati$^{1,3}$
\\ \\ 
$^1$Department of Mathematics and Statistics, 
McGill University, \\
805 Sherbrooke Street West, 
Montreal,  QC,  H3A 2K6, Canada
\\ \\
$^2$Department of Mathematics, Harvard University,\\
One Oxford St., Cambridge, MA, 02138, USA
\\ \\
$^3$Aix-Marseille Universit\'e, CNRS, CPT, UMR 7332, Case 907, 13288 Marseille, France\\
Universit\'e de Toulon, CNRS, CPT, UMR 7332, 83957 La Garde, France\\
FRUMAM
}
\maketitle
{\small
{\bf Abstract.} The property that a Jacobi matrix is reflectionless  is usually characterized either in   terms of Weyl $m$-functions 
or the vanishing of the real part of the boundary values of the diagonal matrix elements of the resolvent. We introduce a
characterization  in terms of stationary scattering theory (the vanishing of the reflection coefficients) and prove that this characterization is 
equivalent  to  the usual ones. We also show that  the new  characterization is equivalent to the notion 
of being dynamically reflectionless,  thus  providing  a short proof of an important  result of \cite{BRS}. The motivation for the new characterization comes from recent studies of the non-equilibrium statistical mechanics of the electronic black box model and we  elaborate on this connection.} 


\thispagestyle{empty}
\section{Introduction}

The purpose of this paper is two fold. First, we wish to advertise a certain point of view regarding full-line Jacobi matrices that appears 
absent from the literature. Secondly, we shall  use this point of view to give a new characterization of   reflectionless  Jacobi 
matrices.

The point of view we wish to describe has its origin in recent studies of the non-equilibrium statistical mechanics of the electronic  black box model. 
We will need here only the  simplest variant of this model. The simple electronic black box (SEBB) model   describes a quantum dot ${\cal S}$ coupled to two electronic reservoirs ${\cal R}_{l/r}$ that we will colloquially call `left' and `right.' 
The system ${\cal S}$ is described by the Hilbert space $\cc$ and energy $\omega \in \rr$. The reservoir ${\cal R}_{l/r}$ is described 
by a pair $(\mfh_{l/r}, h_{l/r})$, where $\mfh_{l/r}$ is a single fermion  Hilbert space and $h_{l/r}$ is a  single fermion Hamiltonian.  We set 
\[\mfh=\mfh_l\oplus \cc \oplus \mfh_r, \qquad h_0= h_l \oplus \omega \oplus h_r.\]
The fermionic second quantization of the pair $(\mfh, h_0)$ describes the uncoupled SEBB  model ${\cal R}_l + {\cal S} + {\cal R}_r$. The junctions coupling 
$\cS$ to the reservoirs are  specified by a choice of unit vectors  $\chi_{l/r}\in \mfh_{l/r}$ and real numbers $\lambda_{l/r}$ describing 
the strength of the coupling.  The corresponding tunnelling Hamiltonian is 
\[h_T=h_{T, l} + h_{T, r}, \qquad  h_{T, l/r}=\lambda_{l/r}\left( \vert1\rangle \langle \chi_{l/r}\vert   +  \vert\chi_{l/r}\rangle \langle 1\vert \right).\]
Finally, the coupled EBB model  is obtained by the fermionic second quantization of the pair 
$(\mfh, h)$ where 
\[ h=h_0 + h_T.\]
In the literature, the Hamiltionian $h$ is also known as the {\em Friedrichs model} or {\em Wigner-Weisskopf atom}. 

In the study of the  SEBB  model it is no loss of generality to assume that 
$\chi_{l/r}$ is a cyclic vector for $h_{l/r}$. We shall further assume that the operators $h_{l/r}$ are bounded. The SEBB model 
is called non-trivial if the spectral measure $\nu_{l/r}$ for $h_{l/r}$ and $\chi_{l/r}$ is not supported on a finite set. 

The relevant  transport phenomena in the SEBB model are linked to  the choice of the initial state. Suppose that  initially 
the reservoir ${\cal R}_{l/r}$ is in equilibrium at inverse temperature $\beta_{l/r}$ and chemical potential $\mu_{l/r}$. If either $\beta_{l}\not=
\beta_r$ or $\mu_l\not=\mu_r$, then, in the large time limit,  the resulting  temperature/chemical potential differential induces a non-trivial  steady state  energy/charge flux between 
the left and the right reservoirs. The Landauer-B\"uttiker formalism  relates 
the   transport theory of the SEBB  model   (the values of the steady state energy and charge fluxes,  full counting statistics of the energy and charge fluxes, etc.) to the scattering theory of the pair $(h, h_0)$. For example, the steady state energy/charge flux is given by 
the Landauer-B\"uttiker formula that involves only the initial energy densities  of the reservoirs and the scattering matrix 
of the pair $(h, h_0)$. The same is true for the  Levitov-Lesovik formulas for  the large deviation  functionals
associated to  the  full counting statistics of 
energy/charge flux. We refer the reader  to the lecture notes \cite{JKP, JOPP}  for additional information  and  references to the vast literature on these subjects.  

Consider now  a full-line Jacobi matrix $J$ acting on $\fh = \ell^2 (\Z)$ by
\beq
(Ju)_k = a_k u_{k+1} + a_{k-1} u_{k-1} + b_k u_k,
\label{jacoby}
\eeq
where $\{a_k \}_{k \in \Z}$ and $\{b_k\}_{k \in \Z}$ are bounded sequences of real numbers and $a_k \neq 0$ for every $k$.
To such a matrix one associates a  SEBB model as follows. For fixed $n \in \zz$,  let  
\[\fhln = \ell^2 ( ( - \infty , n-1  ] ), \qquad \fhrn = \ell^2 ( [ n+1, \infty)).\]
Let  $\Jln$ and $\Jrn$ be the half-line Jacobi matrices with Dirichlet boundary conditions obtained by restricting $J$ to $\fhln$ and $\fhrn$, respectively. Setting $\mfh_{l/r}=\fh_n^{(l/r)}$, $h_{l/r}=J_{n}^{(l/r)}$,  $\omega=b_n$, $\chi_l=\delta_{n-1}$, $\chi_r=\delta_{n+1}$, 
$\lambda_l= a_{n-1}$, $\lambda_r=a_n$,  and
\beq
J_0 = J^{(l)}_n + J^{(r)}_n + b_n \vert \delta_n \rangle \langle \delta_n \vert,
\label{jacoby0}
\eeq
one arrives at the SEBB model with 
\[h_0= J_0, \qquad h= J.
\]
Thus, a Jacobi matrix naturally induces  a non-trivial SEBB model. In turn, 
a  well-known  orthogonal polynomial construction (see Theorem I.2.4  in \cite{Si}) implies that {\em every} non-trivial SEBB model is unitarily equivalent 
to a Jacobi matrix SEBB model. This leads to the  identification of  the class  of   SEBB models with the class of  Jacobi matrices. 

This note is a first step in the exploration of this connection. We shall focus on the following point. Although the study of the scattering theory of the pair $(h, h_0)$ is completely natural from 
the point of view of non-equilibrium statistical mechanics, the study of the scattering theory of the corresponding pair of 
Jacobi matrices $(J, J_0)$  is, to the best of our knowledge, virtually absent in the  literature on Jacobi matrices. We shall exploit this connection to determine when a Jacobi matrix is reflectionless, a property which has attracted  considerable attention in the recent literature on Jacobi matrices (see 
Chapter 8   in \cite{Te} and Chapter 7 in \cite{Si}). Motivated by recent studies of the SEBB model (and in particular  by \cite{JLP}) we shall propose a definition of reflectionless based on the scattering matrix of the pair 
$(J, J_0)$ that is, in our opinion, physically and mathematically natural. We   show that this definition is  equivalent to the  standard definitions appearing
in the literature and also to the dynamical definition introduced in \cite{BRS}. A consequence is 
 a short, transparent proof of the main result of  \cite{BRS}  which has settled a  25 year old conjecture raised in \cite{DeSi}. 
 
 The paper is organized as follows. In Section 2 we recall the standard definitions of reflectionless Jacobi matrices and the result of 
 \cite{BRS}. In Section 3 we introduce stationary reflectionless Jacobi matrices and state our main result.
Its proof is given in 
 Section 4. The concepts introduced in this paper go beyond reflectionless and shed light on the notions of the dynamical reflection 
 probability of \cite{DS} and the spectral reflection probability of \cite{BRS, GNP,  GS} by identifying them with the scattering matrix reflection probability. 
 This point is discussed in Section 5. 
 
 Finally, we remark that, as in  \cite{BRS},  our results can be extended  to  Schr\"odinger operators on the real line and CMV matrices. 
The extension to Schr\"odinger operators is somewhat technical due to the lack of convenient references to the stationary scattering theory 
of singular perturbations (induced by imposing boundary conditions). The details can be found in \cite{Z}.  The extension to CMV matrices can be found in \cite{CLP}.

\bigskip

{\noindent\bf Acknowledgment.}  The research of V.J. and B.L. was partly supported by NSERC. The research of A.P. 
was partly supported by  ANR (grant 12-JS01-0008-01). We wish to thank  J. Breuer, R. Frank and B. Simon for useful discussions. 
A part of this work was done during the visit of the authors to Institute Henri Poincar\'e in Paris. We wish to thank M. Esteban and 
M. Lewin for their hospitality.

\section{Three notions of reflectionless}
The Green's function of the Jacobi matrix  (\ref{jacoby})  is defined for $z \in \cc \backslash \rr$ by
\[
G_{nm} (z) = \langle \delta_n, (J - z)^{-1} \delta_m \rangle.
\]
By general principles (see \cite{J} or any book on harmonic analysis)  the boundary values
\beq
G_{nm} (\lambda \pm \i 0) = \lim_{\eps \dto 0} \langle \delta_n , ( J - \lambda \mp \i \eps )^{-1} \delta_m \rangle \label{eqn:gnmbv}
\eeq
exist, are finite and non-vanishing  for Lebesgue a.e. $\lambda \in \rr$.

The Weyl $m$-functions are defined for $z \in \cc \backslash \rr$ by
\[
\mln (z) = \langle \delta_{n-1} , ( \Jln - z )^{-1} \delta_{n-1} \rangle , \qquad \mrn (z) = \langle \delta_{n+1} , (\Jrn - z)^{-1} \delta_{n+1} \rangle.
\]
The boundary values $\mlrn( \lambda \pm \i 0)$ are defined analogously to (\ref{eqn:gnmbv}), and are also  finite and non-vanishing for Lebesgue a.e. $\lambda \in \rr$. 

We recall the terminology of 
\cite{BRS}. Let $\ee \subset \rr$ be a Borel set. A  Jacobi matrix is called \emph{measure theoretically reflectionless} on  $\ee$ if
\beq
\Re [ G_{nn} ( \lambda + \i 0) ] = 0 \label{eqn:regnn}
\eeq
for Lebesgue a.e. $\lambda \in \ee$ and every $n$.  A  Jacobi matrix is called \emph{spectrally reflectionless} on  $\ee$ if 
\beq
a_n^2 \mrn ( \lambda + \i 0) \mlnp ( \lambda - \i 0) = 1 \label{eqn:specref}
\eeq
for Lebesgue a.e. $\lambda \in \ee$ and every $n$.

From the first of the two formulas
\beq
G_{nn} (z) = - \frac{1}{ a_n^2 \mrn(z) - \mlnp (z) ^{-1} } = - \frac{1}{ a_{n-1}^2 \mln (z) - \mrnm (z)^{-1}}, \label{eqn:gnnform}
\eeq
one derives  that if (\ref{eqn:specref})
holds for $\lambda$ and $n$, then (\ref{eqn:regnn}) holds for the same $\lambda$ and $n$. The following is the well-known converse (see 
\cite{GKT, SY, Te} and Theorem 7.4.1  in \cite{Si}).
\bet \label{thm:equivrefl}For Lebesgue a.e. $\lambda \in \rr$, the following are equivalent.
\ben 
\item $\Re [ G_{nn} ( \lambda + \i 0) ] = 0$ for all $n$.
\item $ \Re [G_{nn} ( \lambda + \i 0) ] = 0$ for three consecutive $n$'s.
\item (\ref{eqn:specref}) holds for one $n$.
\item (\ref{eqn:specref}) holds for every $n$.
\een
\eet
An immediate and well-known consequence is 
\begin{corollary} $J$ is   measure theoretically reflectionless on $\ee$ iff it is  spectrally reflectionless on $\ee$. 
\end{corollary}

Based  on the ideas of Davies and Simon \cite{DS}, Breuer, Ryckman and Simon introduced in  \cite{BRS}  the notion of being dynamically reflectionless. We first recall 
the setup of \cite{DS} adapted to the Jacobi matrix case.  For each $n \in \Z$, let $\cln$ be the characteristic function of $( - \infty, n-1]$ and $\crn$ of $[n+1, \infty)$. Set
\[
\fh_l^{\pm} = \left\{ \varphi \in \fh_{\ac} (J) \,\big|\,  \mbox{ for all } n, \lim_{t \to \pm \infty}  \|\crn \e^{-\i t J} \varphi \| = 0 \right\}.
\]
The elements of $\fh_l^{\pm}$ are the states that are concentrated on the left in the distant future/past  $t \rightarrow \pm \infty$. 
The sets $\fh_r^\pm$ are defined with $\cln$ replacing $\crn$. The sets $\fh_{l/r}^+$ are related to $\fh_{l/r}^-$ by  time-reversal, 
{\sl i.e.}, 
\beq \fh_{l/r}^+=\left\{\bar \varphi \,|\, \varphi \in \fh_{l/r}^-\right\}. \label{eqn:tri}
\eeq

In what follows $\fh_{\ac}(A)$ denotes the absolutely continuous subspace for self-adjoint $A$ and $P_{\ac}(A)$  the projection onto this subspace.
The result of Davies and Simon \cite{DS}  is: 
\bet 
\beq
\fh_{\ac} (J) = \fh_l^+ \oplus \fh_r^+ = \fh_l^- \oplus \fh_r^- .
\label{tri}
\eeq
\label{davies-simon}
\eet
{\bf Remark.} In \cite{DS}  this theorem was proven  in the context of Schr\"odinger operators on the real line. In the discrete 
setting considered here the argument is considerably simpler. For completeness and later reference,  we sketch the proof.  

\proof  By (\ref{eqn:tri})  it suffices to prove that $\fh_{\ac} (J) = \fh_l^+ \oplus \fh_r^+$.  Since 
$\slim_{ t \to  \infty} \e^{ \i t J} C \e^{ - \i t J} P_{\ac} (J) = 0$ for any compact $C$, we have that  for any  $n$ 
\[ 
\fh_{l/r}^{+} = \left\{ \varphi \in \fh_{\ac} (J) \,\big|\,  \lim_{t \to  \infty}  \|\chi^{(r/l)}_n\e^{-\i t J} \varphi \| = 0 \right\}= 
\left\{ \varphi \in \fh_{\ac} (J) \,\big|\,  \lim_{t \to \infty}  \e^{\i t J}\chi^{(l/r)}_n\e^{-\i t J} \varphi =\varphi\right\}. 
\]
Since the commutators $[\chi_n^{(l/r)}, J]$ are finite rank, trace-class scattering theory implies that the limits
\[
P_{l/r}^+= \slim_{t \to  \infty} \e^{ \i t J} \chi_n^{(l/r)} \e^{ - \i t J} P_{\ac} (J) \]
exist, and are furthermore independent of $n$. From the fact that $\e^{ \i t J}$ commutes with $P_{l/r}^+$, we obtain that $P_{\ac}(J)$ commutes with $P_{l/r}^+$, and so
\begin{align*}
[{P_{l/r}^+}]^* &= \left[ \slim_{t \to  \infty} \e^{ \i t J} P_{\ac}(J) \chi_n^{(l/r)} P_{\ac} (J) \e^{- \i t J} \right]^* \\
&= \slim_{t \to  \infty} \e^{ \i t J} P_{\ac} (J) \chi_{n}^{(l/r)} P_{\ac}(J) \e^{- \i t J} = P_{l/r}^+,
\end{align*}
and
\begin{align*}
[{P_{l/r}^+ }]^2 &= \slim_{t \to  \infty} \e^{ \i t J } \ch0lr \e^{ - \i t J} P_{\ac} (J) \e^{ \i t J } \ch0lr \e^{ - \i t J} P_{\ac} (J) \\
&= \slim_{t \to  \infty} \e^{ \i t J } \ch0lr \e^{ - \i t J}  \e^{ \i t J } \ch0lr \e^{ - \i t J} P_{\ac} (J) = P_{l/r}^+.
\end{align*}
Hence,  $P_{l/r}^+$ are  the orthogonal projections onto $\fh_{l/r}^+$. The statement now follows from the identity 
\[
P_{\ac} (J) = \e^{ \i t J} \chi_n^{(l)} \e^{ - \i t J} P_{\ac} (J) +  \e^{ \i t J} \chi_n^{({r})} \e^{ - \i t J} P_{\ac} (J) +
 \e^{ \i t J} |\delta_n\rangle \langle \delta_n| \e^{ - \i t J} P_{\ac} (J)  .\]
 \qed

In what follows $P_\ee (A)$ denotes the spectral projection of a self-adjoint operator 
$A$ onto a Borel set $\ee$ and $|S|$  the Lebesgue measure of a set $S$. In \cite{BRS} the following notion was 
introduced. 
\begin{definition} 
A Jacobi matrix is called dynamically reflectionless on $\ee$ if for any Borel  set $\ee_1 \subset \ee$,
\beq
P_{\ee_1} (J) P_\ac (J) = 0 \implies | \ee_1 | = 0 \label{eqn:dynref},
\eeq
and
\[
P_\ee (J) [ \fh_l^+ ] = P_\ee (J) [ \fh_r^- ].
\]
\end{definition}
The result of \cite{BRS} is
\bet \label{thm:main}
J is dynamically reflectionless on $\ee$ iff it is spectrally reflectionless on $\ee$.
\eet
{\bf Remark.} Theorem \ref{thm:main} verifies the conjecture of  Deift and Simon \cite{DeSi} that the a.c. spectrum of ergodic Jacobi matrices 
is dynamically reflectionless. 

\section{The notion of stationary reflectionless}

Recall that $J_0$ is given by (\ref{jacoby0}). 
 It follows from trace-class scattering theory that the wave operators
\[
w_\pm = \slim_{t \to \pm \infty} \e^{ \i t J} \e^{ - \i t J_0} P_{\ac} (J_0 )
\]
exist and are complete (here, completeness means that $\Ran w_\pm = \fh_{\ac} (J)$). Note that  $w_{\pm}$ depends on $n$ (whenever this dependence is 
clear within the context, we  shall not indicate it  explicitly). By the spectral theorem we may identify $\fh_{\ac} (J_0)$ with
\[
\fh_{\ac}(J_0) = \fh_{\ac} (J^{(l)}_n ) \oplus \fh_{\ac} (J^{({r})}_n) = L^2 ( \rr, \dnl ) \oplus L^2 (\rr, \dnr )
\]
where $\nu_{l/r, \ac}$ is the a.c. part of the spectral measure for $J_0^{(l/r)}$ and $\delta_{n-1} / \delta_{n+1}$. We recall the well-known formula
\[
\frac{ \d \nu_{l/r, \ac} }{\d \lambda } ( \lambda ) = \frac{1}{\pi} \Im [m_n^{(l/r)} ( \lambda + \i 0) ]
\]
(a pedagogical proof can be found in \cite{J}). It follows from stationary scattering theory \cite{Y} that the scattering matrix
\[
s = w_+^* w_-
\]
acts as multiplication by a unitary $2 \times 2$ matrix 
\[
s ( \lambda ) = \left( \begin{matrix} s_{ll} ( \lambda ) & s_{lr} ( \lambda ) \\ s_{rl} ( \lambda ) & s_{rr} ( \lambda ) \end{matrix} \right)
\]
on $\fh_{\ac} (J_0)$ where
\beq
s_{jk} ( \lambda) = \delta_{j, k} + 2 \i a_j a_k G_{nn} ( \lambda + \i 0) \sqrt{ \Im [ m^{(j)}_n ( \lambda + \i 0)] \Im [m^{(k)}_n ( \lambda + \i 0)]}, \label{eqn:scatform}
\eeq
 $j, k \in \{l, r\}$, $a_l = a_{n-1}$ and $a_r = a_n$. In the current setting, the above formula can be also easily  verified by a direct computation which we sketch in the appendix (see \cite{La} for more details). Note that $s_{lr}(\lambda)=s_{rl}(\lambda)$. 
 
 Motivated by the non-equilibrium statistical mechanics of the SEBB model 
(and  in particular by the work \cite{JLP}) we introduce

\begin{definition} A Jacobi matrix $J$ is   called stationary reflectionless on a Borel set  $\ee$ if for one $n$  the scattering 
matrix $s(\lambda)$ is off-diagonal for Lebesgue a.e. $\lambda \in \ee$. 
\end{definition}
{\bf Remark.} In other words, $J$ is stationary reflectionless on $\ee$ if, for one $n$, $|s_{lr}(\lambda)|=1$ for Lebesgue a.e. $\lambda \in \ee$, 
or equivalently, $s_{ll}(\lambda)=s_{rr}(\lambda)=0$ for Lebesgue  a.e. $\lambda \in \ee$.

Formulas (\ref{eqn:gnnform}),  (\ref{eqn:scatform}), and Theorem \ref{thm:equivrefl} immediately give:
\bep $J$ is spectrally reflectionless on $\ee$ iff $J$ is stationary reflectionless on $\ee$.
\label{when}
\eep
{\bf Remark.}  In particular, this proposition implies that  if the scattering 
matrix $s(\lambda)$ is off-diagonal for Lebesgue a.e. $\lambda \in \ee$ and some $n$,  then it is so for all $n$. 

We shall prove 
\bet\label{main-us}
J is dynamically reflectionless on $\ee$ iff it is stationary reflectionless on $\ee$.
\eet
This result combined with Proposition \ref{when} implies Theorem \ref{thm:main}. 

The proof of Proposition \ref{when} is very simple. 
As we shall see,  the proof of Theorem \ref{main-us} is also simple due to the 
direct connection  with scattering theory. The notion of stationary reflectionless naturally links the notions of 
spectral and dynamical reflectionless and is 
likely to play a role in future developments.
\section{Proof of Theorem \ref{main-us}}
Let 
\[
\Sigma_{l/r, \ac} = \left\{ \lambda\,\, \big|\,\, \frac{ \d \nu_{l/r, \ac} } { \d \lambda } ( \lambda ) > 0 \right\} , \qquad \E = \Sigma_{l, \ac} \cup \Sigma_{r, \ac}.
\]
 The set $\E$ is an essential support of the a.c. spectrum of $J_0$. From unitarity and symmetry of the scattering matrix, we see that for Lebesgue a.e. $\lambda$, 
\[
s_{ll}(\lambda) = 0 \iff s_{rr} ( \lambda) = 0.
\]
Note also that if $J$ is stationary reflectionless on $\ee$,  then $|\ee  \backslash ( \Sigma_{l, \ac} \cap \Sigma_{r, \ac} )| =0$, and 
in particular $| \ee \backslash \E | = 0$. These two observations yield an equivalent formulation of stationary reflectionless that 
is more suitable  for comparison with being dynamically reflectionless:
\bel  $J$ is stationary reflectionless on $\ee$ iff  $| \ee \backslash \E |=0$  and, for some $n$, 
\[
s_{ll}( \lambda ) = 0 \mbox{ for Lebesgue a.e. } \lambda \in \ee \cap \Sigma_{l, \ac} , \quad s_{rr} ( \lambda) = 0 \mbox { for Lebesgue a.e. } \lambda \in \ee \cap \Sigma_{r, \ac}.
\]
\label{lemma-fi}
\eel
Since $J-J_0$ is finite rank,  $\E$ is an essential support of the a.c. spectrum of $J$, and so the condition (\ref{eqn:dynref}) in the definition of dynamical reflectionless can be replaced by the equivalent condition
\beq
| \ee \backslash \E | = 0. \label{eqn:dynreform}
\eeq
The key observation is that the projections $P_{l/r}^\pm$ on the subspaces $\fh_{l/r}^\pm$ (recall the proof of Theorem \ref{davies-simon}) satisfy 
\beq
P_{l/r}^\pm = \slim_{t \to \pm \infty} \e^{ \i t J} \e^{ - \i t J_0 } \chi_n^{(l/r)}\e^{ \i t J_0 } e^{ - \i t J} P_{\ac} (J) = w_\pm \chi_n^{(l/r)} w_\pm^*.
\label{dasi-proj}
\eeq
Above, we have used the fact that $J_0$ commutes with $\chi_n^{(l/r)}$ and that $\Ran w_\pm^* = \fh_{\ac} (J_0)$. Note that the $P_{l/r}^\pm$ commute with $J$ and its spectral projections. For $\varphi , \psi \in \fh$,
\begin{align}
\langle \varphi , P_\ee (J) P_l^+  P_l ^- \psi \rangle &=\langle \varphi , P_\ee (J) P_l^+ P_\ee (J) P_l ^- \psi \rangle = \langle w_+ ^* P_\ee (J) \varphi , \chi_n^{(l)} w_+^* w_-  \chi_n^{(l)} w_-^* P_\ee (J) \psi \rangle \notag \\
&= \langle w_+^* \varphi , P_\ee (J_0) \chi_n^{(l)} s \chi_n^{(l)} P_\ee (J_0 ) w_-^* \psi \rangle , \label{eqn:computation}
\end{align}
where the last line follows by the intertwining property of the wave operators. We write
\[
w_+^* \varphi = \widetilde{ \varphi}_l (\lambda) \oplus \widetilde{ \varphi }_r ( \lambda ) \in L^2 (\rr , \dnl ) \oplus L^2 (\rr, \dnr ) ,
\]
and the same for $w_-^* \psi$. With this notation, the last line of (\ref{eqn:computation}) becomes
\[
\int_\ee \bar{\widetilde{\varphi}_l} ( \lambda ) s_{ll} ( \lambda ) \widetilde{\psi}_l (\lambda ) \dnl (\lambda).
\]
Since $\Ran w_\pm^* = \fh_{\ac} (J_0)$, we conclude
\[
P_\ee (J) P_l^+  P_l^- = 0 \iff s_{ll} ( \lambda ) = 0 \mbox{ for Lebesgue a.e. } \lambda \in \ee \cap \Sigma_{l, \ac}.
\]
A similar computation yields
\[
P_\ee (J) P_l^-  P_l^+ = 0 \iff \bar{s}_{ll} ( \lambda ) = 0 \mbox{ for Lebesgue a.e. } \lambda \in \ee \cap \Sigma_{l, \ac},
\]
and two similar statements where $P_{l}^\pm$ is replaced by $P_{r}^\pm$ and $s_{ll}$ by $s_{rr}$. This, together with Lemma 
\ref{lemma-fi}  and (\ref{eqn:dynreform}), yields the theorem.

\section{Remarks}

The arguments used in the proof of Theorem \ref{main-us} go beyond reflectionless and shed  light on the notions of 
dynamical and spectral reflection probability. 

{\em Dynamical reflection probability.} In \cite{DS} (see also \cite{BRS}) Davies and Simon introduced the concept of dynamical reflection probability (also called reflection 
modulus) as follows. Recall that the projections $P_{l/r}^\pm$ satisfy (\ref{dasi-proj}). $J$ commutes with 
$P_l^{+}P_l^{-}P_l^{+}$ and, since $J\upharpoonright \Ran P_l^+$ has simple spectrum, there exists a Borel function $R_l^+$ on 
$\sp(J\upharpoonright \Ran P_l^{+})$ such that 
\[P_l^{+}P_l^{-}P_l^{+}=R_l^+(J)\upharpoonright \Ran P_l^+.\]
$R_l^+$ is unique (up to sets of Lebesgue measure zero) and satisfies $0\leq R_l^+(\lambda)\leq 1$. One extends $R_l^+$ to $\rr$ by setting 
$R_l^+(\lambda)=1$ for $\lambda \not\in \sp(J\upharpoonright \Ran P_l^{+})$ and defines $R_l^{-}$, $R_r^{\pm}$ analogously. The functions 
$R_{l/r}^{\pm}$ are discused in Section 4 of \cite{DS} in the Schr\"odinger case (their proofs  easily extend  to  the Jacobi case). In this 
context, the main observation of this note is that the formula (\ref{dasi-proj}) implies the identity 
\[ R_{l/r}^\pm(\lambda)=|s_{ll}(\lambda)|^2=|s_{rr}(\lambda)|^2.\]

{\em Spectral reflection probability.}
To the best of our knowledge, the link between reflection probability and half-line $m$-functions was first observed in \cite{GNP, GS} in the context 
of  Schr\"odinger operators on the line. The definition of the spectral reflection probability of \cite{GNP, GS} was based on suitable generalized eigenfunctions and was 
extended to Jacobi matrices in \cite{BRS} as follows.  Consider the case $n=0$. For $z\in \cc_+$, let 
$\psi^{l/r}(z) =\{\psi_k^{(l/r)}(z)\}_{k\in \zz}$ be the unique solution of the equation 
\beq a_k\psi_{k+1} + a_{k-1}\psi_{k-1} + b_k\psi_k = z\psi_k
\label{voi}
\eeq
that is square summable at $\mp \infty$ and normalized by $\psi_{0}^{(l/r)}=1$. These solutions are related to $m$-functions as
\beq
m_0^{({r})}(z)=-\frac{\psi_1^{({r})}(z)}{a_0}, \qquad m_1^{(l)}(z)=-\frac{1}{a_0\psi_1^{(l)}(z)}.
\label{m-psi}
\eeq
For all $k$ and Lebesgue a.e. $\lambda$ the limit 
\[\lim_{\epsilon \downarrow 0}\psi_k^{(l/r)}(\lambda +\i \epsilon)=\psi_k^{(l/r)}(\lambda +\i 0)\]
exists and  $\psi^{(l/r)}(\lambda +\i 0)$ solves (\ref{voi}) with $z=\lambda$. 
For Lebesgue a.e. $\lambda \in \Sigma_{r, {\rm ac}}$ the solution $\psi^{({r})}(\lambda +\i 0)$ is not a multiple of 
a real solution and so $\bar {\psi^{({r})}(\lambda +\i 0)}$ is also a solution linearly independent of $\psi^{({r})}(\lambda +\i 0)$. Hence, 
for Lebesgue a.e. $\lambda \in \Sigma_{r, {\rm ac}}$ we can expand 
\beq \psi^{({l})}(\lambda +\i 0)= \alpha(\lambda)\bar{\psi^{({r})}(\lambda +\i 0)} +\beta(\lambda)\psi^{({r})}(\lambda +\i 0).
\label{expa}
\eeq
The spectral reflection probability  of \cite{GNP, GS, BRS} is 
\[ {\cal R}_r(\lambda)=\left|\frac{\beta(\lambda)}{\alpha(\lambda)}\right|^2.
\]
One extends ${\cal R}_r(\lambda)$ to $\rr$ by setting ${\cal R}_r(\lambda)=1$ for $\lambda \not \in \Sigma_{r, {\rm ac}}$ and 
defines ${\cal R}_l(\lambda)$ analogously. Using (\ref{m-psi}) and  (\ref{expa}) one computes 
\[ |{\cal R}_r(\lambda)|^2=\left|\frac{ a_0^2\bar{m_0^{({r})}(\lambda +\i 0)} m_1^{(l)}(\lambda +\i 0) -1}
{ a_0^2{m_0^{({r})}(\lambda +\i 0)} m_1^{(l)}(\lambda +\i 0) -1}\right|^2.
\]
In this context the  main observation of this  note is that the formulas  (\ref{eqn:gnnform})  and (\ref{eqn:scatform}) yield  the identity 
\[
{\cal R}_{r}(\lambda)=|s_{rr}(\lambda)|^2.
\]
Similarly, 
\[
{\cal R}_{l}(\lambda)=|s_{ll}(\lambda)|^2.
\]
These identities   clarify the meaning of the spectral reflection probability and yield
\beq 
{\cal R}_{l/r}=R_{l/r}^\pm.
\label{BRS} \eeq
Much of the technical 
work in \cite{BRS} was devoted to a direct proof of (\ref{BRS}) via an implicit rederivation of the scattering matrix.

\appendix
\section{Computation of the scattering matrix}

We briefly sketch the derivation of the formula (\ref{eqn:scatform}) for the scattering  matrix of the 
pair $(J, J_0)$.  For details, we refer the reader to \cite{La}.

It suffices to consider the case $n=0$  (recall \ref{jacoby}). 
First,  we shall show that  for $\varphi \in \fh$,
\[
w_\pm^* \varphi = \varphi^{(\pm)}_l \oplus \varphi^{(\pm)}_r \in \fh_{\ac} ( \Jl0 ) \oplus \fh_{\ac} ( \Jr0 ),
\]
where (recalling that $a_l = a_{-1}$ and $a_r = a_0$),
\[
\varphi_{l/r} ^{(\pm)} (\lambda) = P_{\ac} (\Jlr0 )  \varphi  (\lambda) - a_{l/r} \langle \de0 , ( J - \lambda \mp \i 0)^{-1} \varphi \rangle.
\]
For any $\psi = \psi_l \oplus \psi_r \in  \fh_{\ac} ( \Jl0 ) \oplus \fh_{\ac} ( \Jr0 )$ we have,
\[
\langle \psi , w_+^* \varphi \rangle = \langle w_+ \psi, \varphi \rangle = \lim_{t \to \infty} \langle \e^{ \i t J } \e^{ - \i t J_0 } \psi , \varphi \rangle = \lim_{t \to \infty} \langle \psi, \e^{ \i t J_0 } \e^{ - \i t J } \varphi \rangle
\]
By an abelian limit  and the definition of $J - J_0$ we can rewrite the RHS as
\[
\lim_{t \to \infty} \langle \psi , \e^{ \i t J_0} \e^{ - \i t J} \varphi \rangle = \langle \psi , \varphi \rangle - \i \lim_{t \to \infty} \int_0^t \langle \psi , \e^{ \i s J_0 } (J - J_0 ) \e^{ - \i s J} \varphi \rangle \d s = \langle \psi , \varphi \rangle - \lim_{\eps \dto 0 } (L_l ( \eps ) + L_r ( \eps )), 
\]
where
\[
L_{l/r} ( \eps ) = \i \int_0 ^ \infty e^{ - \eps s} a_{l/r} \langle \psi , \e^{ \i s J_0 } \dlr \rangle \langle \de0 ,\e^{ - \i s J } \varphi \rangle \d s .
\]
Expanding the first inner product in the above integrand yields
\begin{align*}
L_{l/r} ( \eps) &=  \i a_{l/r} \int_{\rr} \bar{\psi}_{l/r} ( \lambda ) \left[ \int_0^\infty \langle \de0 , \e^{ - \i s ( J - \lambda - \i \eps ) } \varphi \rangle \d s \right] \dnlr ( \lambda ) \\
&=  a_{l/r} \int_\rr \bar{\psi}_{l/r} ( \lambda ) \langle \de0 , ( J - \lambda - \i \eps )^{-1} \varphi \rangle \dnlr ( \lambda ).
\end{align*}
For $\psi$ in a judiciously chosen dense set (see \cite{La} or Proposition 7 in \cite{JKP})  one can  take the limit $\eps \to 0^+$ inside the integral. This yields the formula for $w_+^*$. The computation for $w_-^*$ is identical.

To compute the scattering matrix, note that for $\psi = \psi_l \oplus \psi_r$ and $\varphi = \varphi_l \oplus \varphi_r$ in $\fh_{\ac} ( J_0 )$,
\begin{align*}
\langle \psi , ( s - \one ) \varphi \rangle &= \langle \psi , ( w_+^* w_- - w_-^* w_- ) \varphi \rangle \\
&= \langle ( w_+ - w_- ) \psi , w_- \varphi \rangle \\
&= \lim_{t \to \infty} \langle ( \e^{ \i t J } \e^{ - \i t J_0 } - \e^{ - \i t J} \e^{ \i t J_0 }) \psi , w_- \varphi \rangle  \\
&= \lim_{t \to \infty} -\i \int_{-t}^t \langle \e^{ \i s J} ( J - J_0 ) \e^{ - \i s J_0 } \psi , w_- \varphi \rangle \d s \\
&= \lim_{ \eps \dto 0} -\i \int_\rr \e^{ - \eps | s | } \langle \e^{ \i s J} (J - J_0) \e^{ - \i s J_0 } \psi , w_- \varphi \rangle \d s.
\end{align*}
The inner product in the above integrand equals
\[
a_l \langle \e^{ - \i s J_0 } \psi , \delta_{-1}  \rangle  \langle w_-^* \de0 , e^{ - \i s J_0 }\varphi \rangle + a_r \langle \e^{ - \i s J_0 } \psi , \delta_{1}  \rangle \langle  w_-^* \de0 , \e ^{ - \i s J_0 } \varphi \rangle,
\]
where  we have used the intertwining property of the wave operators. We use our formula for $w_-^*$ to compute
\[
\langle \psi , (s - \one ) \varphi \rangle = \lim_{\eps \dto 0} \i ( H_{ll} ( \eps) +  H_{rl} ( \eps)  +  H_{lr} ( \eps)  +  H_{rr} ( \eps)  ),
\]
where
\[
H_{jk} ( \eps ) = a_j a_k \int_\rr \e^{ - \eps |s| } \left[  \int_\rr \e ^{ \i s \lambda } \bar{\psi}_j ( \lambda ) \dnj ( \lambda ) \right] \left[ \int_\rr \e^{ - \i s \lambda ' } G_{00} ( \lambda ' + \i 0) \varphi_k ( \lambda' )  \dnk ( \lambda') \right] \d s ,
\]
for $j, k \in \{ l, r \}$. Formally, the computation is completed by noting that 
\[
H_{jk} ( \eps ) = a_j a_k \int_\rr \int_\rr \bar{\psi}_j ( \lambda ) \varphi_k ( \lambda') G_{00} ( \lambda' + \i 0) \left[ \int_\rr \e^{ \i s ( \lambda - \lambda ' ) - \eps |s | } \d s \right] \dnj ( \lambda) \dnk ( \lambda ') ,
\]
and that 
\[ 
\int_\rr e^{ \i s ( \lambda - \lambda') - \eps | s| } \d s \to 2 \pi \delta ( \lambda - \lambda ')
\]
as $ \eps \to 0^+$. For a suitable dense set of $\psi$ and $\varphi$, this formal computation can be easily  justified (see 
\cite{La}).

Finally, we remark that the scattering matrix formula (\ref{eqn:scatform})  is valid only after a `unitarity' transformation which we describe now. For any $\psi \in \fh_{\ac} (J_0 )$, let $\psi ( \lambda )$ denote the vector $( \psi_l ( \lambda ) , \psi_r ( \lambda ) ) \in \cc^2$. For $\psi$ and $\varphi$ in $\fh_{\ac} (J_0)$, we have
\[
\langle \psi , \varphi \rangle = \int_\rr \langle V(\lambda ) \psi ( \lambda ) , V ( \lambda ) \varphi ( \lambda ) \rangle_2 \d \lambda
\]
where $\langle \cdot , \cdot \rangle_2$ denotes the standard inner product on $\cc^2$ and $V ( \lambda )$ is the $2 \times 2$ matrix
\[
V ( \lambda ) = \left( \begin{matrix} \sqrt{ \frac{ \dnl } {\d \lambda }  ( \lambda ) } & 0 \\ 0 & \sqrt{ \frac{ \dnr } { \d \lambda } ( \lambda ) } \end{matrix} \right) .
\]
Multiplication by the matrix $V( \lambda )$ is a unitary operator $V : \fh_{\ac} (J_0 ) \to L^2 ( \rr , \rho_l (\lambda ) \d \lambda ) \oplus L^2 ( \rr , \rho_r ( \lambda ) \d \lambda ) $, with $\rho_{l/r}$ the characteristic function of $\Sigma_{l/r, \ac}$. Our computation shows that the operator  $V s V^{-1}$ acts as multiplication by the $2 \times 2$ matrix $s ( \lambda )$ given by (\ref{eqn:scatform}) on $V \fh_{\ac} (J_0)$. In particular, this transformation ensures that $s ( \lambda)$ is unitary w.r.t. the standard inner product on $\cc^2$. 



\begin{thebibliography}{99999999}



\bibitem[BRS]{BRS} Breuer, J., Ryckman, E., Simon, B.: Equality of the spectral and dynamical definitions of reflection. Commun. Math. Phys. {\bf 295}, 531-550 (2010).

\bibitem[CLP]{CLP} Chu, S., Landon, B., Panangaden, J.: In preparation.

\bibitem[DaSi]{DS} Davies, E.B., Simon, B.: Scattering theory for systems with different spatial asymptotics on the left and right. Commun. Math. Phys. {\bf 63}, 277-301 (1978).

\bibitem[DeSi]{DeSi} Deift, P., Simon, B.: Almost periodic Schr\"odinger operators III. The absolutely continuous spectrum in one dimension. 
Commun. Math. Phys. {\bf 90}, 398-411 (1983).

\bibitem[GKT]{GKT} Gesztesy, F., Krishna, M., Teschl, G.: On isospectral sets of Jacobi operators. Commun. Math. Phys. {\bf 181}, 631-645 (1996).

\bibitem[GNP]{GNP} Gesztesy, F., Nowell, R., Potz, W. One-dimensional scattering theory for
quantum systems with nontrivial spatial asymptotics. Diff. Integral Eqs. {\bf 10},
521-546 (1997).

\bibitem[GS]{GS} Gesztesy, F., Simon, B.: Inverse spectral analysis with partial information on the potential, I. The case of an a.c. component 
in the spectrum. Helv. Phys. Acta {\bf 70}, 66-71 (1997).

\bibitem[J]{J} Jak\v si\'c, V.: Topics in spectral theory. In {\it Open Quantum Systems I. The
Hamiltonian Approach.} S.~Attal, A.~Joye and C.-A.~Pillet editors. Lecture Notes in Mathematics
{\bf 1880}, Springer, Berlin,  2006.

\bibitem[JKP]{JKP} Jak\v si\'c, V., Kritchevski, E., Pillet, C.-A.: Mathematical theory of the
Wigner-Weisskopf atom. In {\em Large Coulomb Systems.} J.~Derezi\'nski and H.~Siedentop editors.
Lecture Notes in Physics {\bf 695}, Springer, Berlin, 2006.

\bibitem[JLP]{JLP} Jak\v{s}i\'c, V., Landon, B., Pillet, C.-A.: Entropic fluctuations of XY quantum spin chains and reflectionless Jacobi matrices. 
Ann. Henri Poincar\'e {\bf 14},  1775-1800 (2013).


\bibitem[JOPP]{JOPP} Jak\v si\'c, V., Ogata, Y., Pautrat, Y., and Pillet, C.-A.:
\newblock Entropic fluctuations in quantum statistical mechanics--an introduction.
\newblock In {\sl Quantum Theory from Small to Large Scales.}
\newblock J.~Fr\"ohlich, M.~Salmhofer, V.~Mastropietro, W.~De Roeck
and L.F.~Cugliandolo editors.
\newblock Oxford University Press, Oxford, 2012.

\bibitem[La]{La} Landon, B.: Master's thesis, McGill University (2013).  





\bibitem[Si]{Si} Simon, B.: {\em  Szeg\"o's  Theorem and its Descendants. Spectral Theory for $L^2$ Perturbations of Orthogonal Polynomials.} M. B. Porter Lectures. Princeton University Press, Princeton, NJ, 2011.

\bibitem[SY]{SY} Sodin, M., Yuditskii, P.: Almost periodic Jacobi matrices with homogeneous spectrum, infinite dimensional Jacobi inversion, and Hardy spaces of character-automorphic functions. J. Geom. Anal. {\bf 7}, 387-435 (1997).



\bibitem[Te]{Te} Teschl, G.: {\em Jacobi Operators and Completely Integrable
Nonlinear Lattices.} Mathematical Surveys and Monographs {\bf 72}, AMS, Providence 2000.




\bibitem[Y]{Y} Yafaev, D.R.: {\em Mathematical Scattering Theory. General Theory.} 
Translated from Russian by J. R. Schulenberger. Translations of Mathematical Monographs, 105. 
American Mathematical Society, Providence, RI, 1992.

\bibitem[Z]{Z} Zwicker, J.: Master's thesis, McGill University, in preparation.

\end{thebibliography}
\end{document}